\documentclass{aastex631}

\graphicspath{{./}{figures/}}
\usepackage{amsmath}
\usepackage{soul}

\begin{document}

\title{Understanding the variability of helium abundance in the solar corona using three-fluid modeling and UV observations}

\author[0000-0003-0602-6693]{Leon Ofman}
\affiliation{NASA Goddard Space Flight Center, Greenbelt, MD, 20771, USA}
\affiliation{The Catholic University of America, Washington, DC 20064, USA}
\affiliation{Visiting, Tel Aviv University, Tel Aviv, Israel}

\author[0000-0001-6018-9018]{Yogesh}
\affiliation{NASA Goddard Space Flight Center, Greenbelt, MD, 20771, USA}
\affiliation{The Catholic University of America, Washington, DC 20064, USA}

\author
[0000-0002-3468-8566]{Silvio Giordano}
\affiliation{INAF – Astrophysical Observatory of Torino, Via Osservatorio 20, 10025 Pino Torinese, Italy}

\begin{abstract}
The variability of helium abundance in the solar corona and the solar wind is an important signature of solar activity, solar cycle, solar wind sources, as well as coronal heating processes. Motivated by recently reported remote sensing UV imaging observations by Helium Resonance Scattering in the Corona and Heliosphere (HERSCHEL) payload sounding rocket of helium abundance in inner corona on 14-Sep-2009 near solar minimum, we present the results of the first three-dimensional three-fluid (electrons, protons, alpha particles) model of tilted coronal streamer belt and slow solar wind that illustrate the various processes leading to helium abundance differentiation and variability. We find good qualitative agreement between the three-fluid model and the coronal helium abundances variability reported from UV observations of streamers, providing insight on the effects of the physical processes, such as heating, gravitational settling and inter-species Coulomb friction in the out-flowing solar wind that produce the observed features. The study impacts our understanding of the origins of the slow solar wind. 
\end{abstract}

\section{Introduction} \label{sec:intro}
The solar wind consists typically of approximately 95\% protons ($H^+$), making them the principal constituent by mass, and around 5\% helium (predominantly $He^{++}$ or $\alpha$ particles) as the second major constituent, and the corresponding number of electrons stemming from charge neutrality. Here, the abundance of helium in the solar wind is defined relative to protons as $A_{He}$ = ($n_{He}/n_p$)$\times$100, where $n_{He}$ and $n_p$ are the concentrations of helium and protons in the solar wind. Understanding variations in helium abundance in the solar wind and solar corona can provide important clues to coronal processes, such as the coronal sources of the slow and fast solar wind, as well as heating and acceleration processes of the solar wind  \citep{Neugebauer1996,Kasper2007,Kasper2012,Abb16}. 

The $A_{He}$ was determined with  remote sensing of UV emission in the inner solar corona by using Sounding-rocket Coronagraph Experiment
(SCORE), the Helium Coronagraph (HeCOR) and 
Extreme Ultraviolet Imaging Telescope (HEIT) in the he Helium Resonance Scattering in the Corona and Heliosphere (HERSCHEL) payload sounding rocket, launched from White Sands, NM on 14-Sep-2009 \citep{Moses2020}. In the photosphere, $A_{He}$ is nearly 8.5\% \citep{Grevesse1998, Asplund2009}. However, it remains about 4–5\% in the solar corona due to the First Ionization Potential (FIP) effects \citep{Laming2001, Laming2003}. The $A_{He}$ can range from 0.1\% to more than 10\% in the solar wind depending on sources, coronal, and interplanetary modulations and due to the heavy mass of $He$ can account for significant fraction of the solar wind mass and energy fluxes \citep{Kasper2007, Alterman2019, Yogesh2021, Yogesh2022, Yogesh2023}. At times, helium can account for the bulk of the solar wind mass flux at the high end limit of $A_{He}$. The long-term variability of $A_{He}$  follows the solar cycle (SC) and varies with solar wind velocity \citep{Kasper2007, Alterman2019, Yogesh2021}. It can increase up to 30\% in coronal mass ejections \citep[][and references therein]{Bor82,Yogesh2022}. Although the above results are mainly observed in the solar wind at 1AU, the variation in helium abundance in the inner solar corona was  studied using remote sensing observations \citep{Moses2020}. Previous 2.5D three-fluid models of the slow solar wind that include electrons, protons, and helium as separate interacting fluids  demonstrate the variability of A(He) associated with streamer structures, heating and coronal sources of the solar wind \citep{Ofman2004,Ofman2004b,Li06,Gio07,OfmKra2010}. These studies modeled the proton and helium streamer structures using 2.5D three-fluid models and predicted their expected signatures long before the feasibility of such UV observations (see, e.g., Figure~4 in \citet{Gio07}). 
Despite these studies, the basic processes that result in the non-uniformity in $A_{He}$ in the solar corona are not well understood in detail so far, due to the complexity of the slow solar wind formation processes \citep[e.g.][]{Abb16}. Here, we use an idealized 3D multi-fluid model to demonstrate the roles of the main ion differentiation physical processes that affect the coronal helium abundance variation in a coronal streamer belt. This is an important problem, as understanding coronal helium abundances  provides important insights into the formation and coronal sources of the slow solar wind.  In Section~\ref{sec:obs} we present the observational motivation, in Section~\ref{sec:model} we present the 3D multi-fluid model, Section~\ref{sec:numresult} is devoted to numerical results, with the discussion and conclusions in Section~\ref{sec:disc}.

\section{Remote sensing observations of coronal helium abundance}\label{sec:obs}
\begin{figure}
\begin{center}
\includegraphics[width=7in]{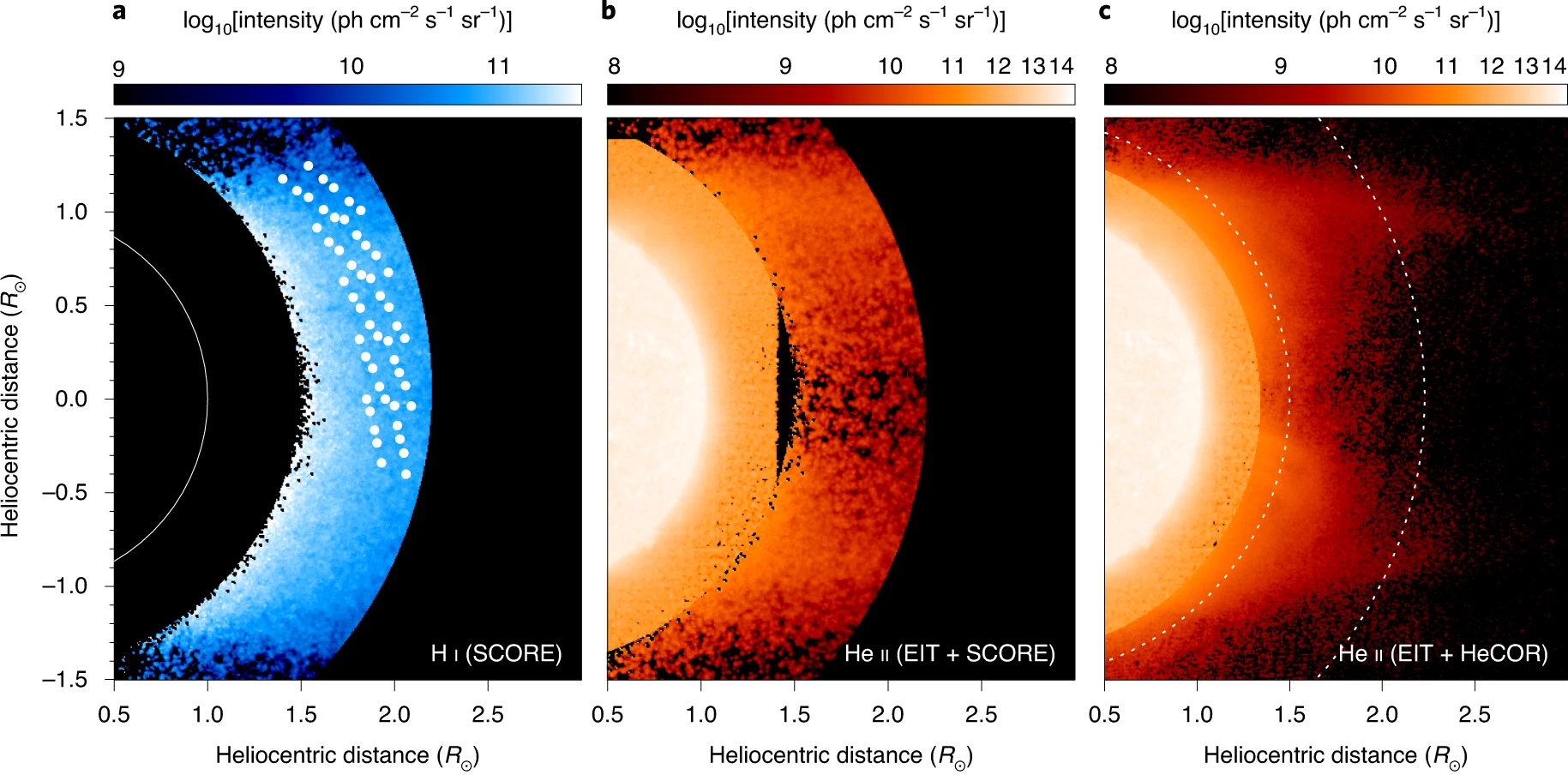}
\caption{Results from the HERSCHEL experiment on 14-Sep-2009. Photometric images were acquired in (a) the SCORE H I $Ly\alpha$, (b) SCORE He~II channels  spanning the range from 1.5$R_s$ to 2.2$R_s$. The white dots in panel (a) denote the positions corresponding to the coordinated SOHO UVCS observations. (c) A photometric image was captured in the HeCOR He~II channel, covering the range from 1.28 to 3$R_s$. The EIT image of the stray light–corrected photometric data of He~II channel are shown in (b) and (c). Reproduced with permission from \citet{Moses2020}.}
\label{fig:herschel}
\end{center}
\end{figure}

Recently, \citet{Moses2020} reported the global helium abundance variability in the solar corona, analyzing the Helium Resonance Scattering in the Corona and Heliosphere (HERSCHEL) sounding rocket observations launched on 14 September 2009, at the extended solar activity minimum of cycle 23. The HERCHEL payload was composed of the Sounding-rocket Coronagraph Experiment (SCORE), the Helium Coronagraph (HeCOR) and the HERSCHEL Extreme Ultraviolet Imaging Telescope (HEIT). SCORE coronagraph is designed to observe the H~I (121.6 nm), He II (30.4 nm) and white light, simultaneously from 1.5 $R_s$ to $2.2 R_s$. HeCOR coronagraph was designed to observe He II (30.4 nm) image from 1.3 $R_s$ to 3 $R_s$ (where $R_s$ is the solar radius). The disk images of He II were  observed by HEIT. The observations from HERCHEL were complemented by Solar and Heliospheric Observatory (SOHO) Extreme ultraviolet Imaging Telescope (EIT), the Large Angle and Spectrometric Coronagraph (LASCO) and the Ultraviolet Coronagraph Spectrometer (UVCS) instruments. In particular, the SCORE was used to obtain the H~I and He~II emission images of the west limb of the solar corona. \citet{Moses2020} found that the morphology of the H~I corona is markedly different from that of the He~II corona, owing to significant spatial variations in helium abundance. Figure~\ref{fig:herschel} shows the observations of the H~I and He~II as reported in \citet{Moses2020}. Figures  \ref{fig:herschel}a and \ref{fig:herschel}b show the images of the H~I and He~II emission respectively, observed by SCORE. Figure~\ref{fig:herschel}b shows the EIT image of the stray light–corrected
photometric He~II channel is superposed inside the inner radius of the SCORE field of view. Figure~\ref{fig:herschel}c shows the He~II emission image from HeCOR superposed with the EIT image inside the HeCOR field of view. Since these observations were focused on the equatorial region in the range of distances from 1.3 to 4$R_s$, the contribution of  other ions such as Si~XI emission to the streamer structure observed in He~II is negligible \citep{Fin03}.

\begin{figure}
\begin{center}
\includegraphics[angle=270,width=4in]{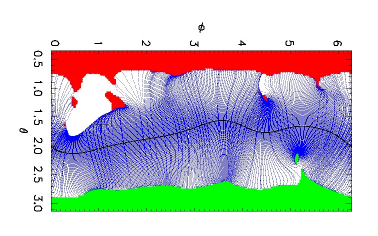}
\caption{Representation of global magnetic field topology from Potential-field Source-surface (PFSS) model \citep{WangSheeley1992} for the Carrington rotation centered on HERSCHEL observation date (model available at NISP NSO Integrated Synoptic Program). The closed field lines are plotted in blue, Regions of open positive and negative flux are represented by green and red, respectively. The neutral line is drawn in black.}
\label{fig:pfss}
\end{center}
\end{figure}

It can be seen from Figure~\ref{fig:herschel} that the H~I emission is nearly uniformly distributed in nearby equatorial regions, whereas the He~II emission is dimmed near equator as compared to higher and lower latitudes. It was found that the corresponding helium abundance was low ($\leq$ 3 \%) within $\pm$  $15^o$ across the equator and the minimum observed $A_{He}$ was 0.6\% near equator. By using potential (`vacuum') magnetic field approximation, \citet{Moses2020} showed that the center of the region exhibiting the lowest $A_{He}$, is situated between two closed-loop systems.  Beyond the latitude range of $\pm15^o$ and beyond the distance of 1.7 $R_s$ where magnetic field lines are mostly open, the relative helium abundance increases. They attributed these changes in helium abundance to the Coulomb drag which depends on the velocity of solar wind, and gravitational settling in the closed field region. In open field line regions, lesser expansion factors cause higher speed resulting in the higher $A_{He}$.  In Figure~\ref{fig:pfss} we show for context the Carrington rotation obtained from NISP NSO Integrated Synoptic Program centered on HERSCHEL observation date with PFSS field where the open and closed  field regions and the tilted streamer belt structure are evident, consistent with the modeling results discussed in Section~\ref{sec:numresult} below. In Figure 2 of \citet{Moses2020} it is apparent that the Helium intensity and abundance latitudinal dependence minima at heights 1.7, 1.9, and 2.1$R_s$ are a few degrees below the 270$^{o}$ latitude. 
The potential magnetic field extrapolation of the double streamer structure presented in Figure~3 of \citet{Moses2020}  is not well consistent with the overall tilted streamer belt structure at solar minimum shown in Figure~\ref{fig:pfss} and the closed magnetic field in the equatorial region. Since the gravitational settling of $He^{++}$ is taking place in closed field of a quiescent streamer and the Coulomb friction enhancement is taking place in open field as demonstrated in past multi-fluid models, the  single streamer belt structure provides more consistent magnetic field structure with the observed helium emission.  

 In their Figure~2 \citet{Moses2020} show the latitudinal variation of the He abundance at the west solar limb derived from SCORE measurements at several heights and found that the He$^{++}$ relative abundance decreases by an order of magnitude at the center of the streamer belt compared to the open field regions, i.e., in the closed field regions in which the plasma is mainly confined. The corresponding variation of H~I in this region was found to be small with overall increase and some dips in intensity. Thus, it is likely that the difference in proton and He$^{++}$ abundance variability arises from ion-dependent physical processes such as gravitational scale-height dependent ($H=k_BT_i/m_ig_s$, where $k_B$ is  Boltzmanns' constant, $T_i$ is ion temperature, $m_i$ is the ion mass, and $g_s$ is the solar gravitational acceleration) settling in the plasma confined to closed magnetic field regions. The observed depletion of helium abundance in the slow solar wind was studied recently using Wind data at 1AU and PSP data at perihelia in \citet{Yogesh2024}, where they suggest that quiescent streamer cores are the preferred locations for gravitational settling to occur. This settling can cause a very low helium abundance, even below 1\% in the solar wind protons. The role of these physical processes is demonstrated using the 3D model three-fluid model in upcoming sections.


\section{Description of the 3D multi-fluid model}
\label{sec:model}
In previous multi-fluid studies, the streamer model incorporated a heavy ion population as an additional fluid alongside plasma electrons and protons \citep[see the review,][]{Abb16}. The plasma is characterized using a three-fluid approximation encompassing electrons, protons, and heavier ions, treated as coupled fluids influenced by collisional and electromagnetic interactions. This approximation extends well beyond the capabilities of single-fluid magnetohydrodynamics (MHD) description. The model assumes quasi-neutrality of the plasma and neglects electron inertia, employing $m_e \ll m_p$ to solve the electron momentum equation for the electric field (i.e., generalized Ohm's law). The electron density is derived from the charge neutrality condition. Notably, viscosity and explicit thermal conduction terms are disregarded in this model. The 2.5D three-fluid model was first developed for streamers with  $O^{5+}$, and $He^{++}$ ions  \citep[e.g.,][]{Ofman2000,Ofman2004,Li06,Ofman2011}. The 3D three-fluid model with $O^{5+}$ ions of a slow solar wind in a tilted streamer belt was presented in \citet{Ofman2015}, while here we include $He^{++}$ ions as the third fluid. The normalized equations employed in the three-fluid model and the detailed description are given in \citet{Ofman2015}. Below, we review the main model parameters relevant to the present study.

 In the model equations shown in \citet{Ofman2015}, the index $k=e$, $p$, $i$ represents electrons, protons, or heavier ions (in the present study $He^{++}$), where the $n_k$ denotes the number density, $V_k$ is the velocity vector, $T_k$ is the temperature for each species, $S_k$ is the empirical heating term (only used for $He^{++}$  ions), $S_{r,e}$ is the electron radiative loss term, $\delta_{k,e}$ is the Kronecker delta, $C_{kjl}$ is the energy coupling term between the species \citep{Ofman2004}, and $\gamma_k$ is the polytropic index of each species. Empirical polytropic indices of $\gamma_k$ = 1.05 are assumed to model the effects of coronal heating and heat conduction loses for electrons, protons  and He$^{++}$ ions   consistent with global coronal models, e.g., \citet{Ril06}. These empirical values of $\gamma_k>1$ allow realistic multifluid modeling of the coronal streamer constituents temperatures in agreement with observations \citep[e.g.,][]{Ofm11}. The terms proportional to ion gyrofrequency are neglected in the low-frequency limit, and the time is normzlized in units of the Alfv\'{e}n time $\tau_A =R_s/V_A$, where the Alfv\'{e}n speed is $V_A = B_0/\sqrt{4\pi m_p n_{e0}}$ defined here with the proton mass, $B_0$ is the normalization magnetic field strength (in the present study we set $B_0$ = 7G), $m_p$ is the proton mass, and $n_{e0}$ is the normalization value of the electron number density. Using $n_{e0} = 5 \times 10^8$ ${cm}^{-3}$, we get $V_A$ = 683 ${km \ s}^{-1}$. For $He^{++}$ we have $A_{He^{++}}=4$, and $Z_{He^{++}}=2$. The other model parameters are $S$ the Lundquist number (here, we set $S=10^4$, typical numerical resolution-limited value in MHD models that does not significantly affect the result); electron and proton Euler number $E u_{\mathrm{e}, \mathrm{p}}=$ $\left(k_{\mathrm{b}} T_{0, \mathrm{e}, \mathrm{p}} / m_{\mathrm{p}}\right) V_{\mathrm{A}}^{-2}$; ion Euler number $E u_{\mathrm{i}}=\left(k_{\mathrm{b}} T_{0, \mathrm{i}} / m_{\mathrm{i}}\right) V_{\mathrm{A}}^{-2}$; and Froude number $F_r=V_{\mathrm{A}}^2 R_{\mathrm{S}} / G M_{\mathrm{S}}$, where $G$ is the universal gravitational constant and $M_{\mathrm{S}}$ is the solar mass, $R_{\mathrm{S}}$ is the solar radius, $b=c B_0 /\left(4 \pi e n_{\mathrm{e} 0} R_{\mathrm{S}} V_{\mathrm{A}}\right)$ the normalization constant, $k_{\mathrm{b}}$ is the Boltzmann constant, and  ${F}_{k{,coul}}$ is the Coulomb friction (or drag) terms proportional to the velocity differences between the species and the collision frequencies \citep[see,e.g.,][for details of these terms]{Braginskii1965, Geiss70,OfmanDavila2001}. The empirical heating term for $\mathrm{He}^{++}$ ions is the function $S_{\mathrm{i}}=S_{\mathrm{i} 0}(r-$ 1) $e^{-(r-1) / \lambda_{\mathrm{i} 0}}$, where the constants $S_{\mathrm{i} 0}$ and $\lambda_{\mathrm{i} 0}$ determine the magnitude and the decrease of the heating rate with distance $r$. In the present model we use the empirical values $S_{\mathrm{i} 0}=7.25$ and $\lambda_{\mathrm{i} 0}=0.5R_s$ constrained by radial density and temperature structure of He in open field regions \citep[e.g.,][]{Ofman2004b}. The initial temperatures were uniform $T_e=T_p=1.6$ MK, $T_{He}=4$ MK, and evolve self-consistently. For simplicity, the possible effects of ion temperature anisotropy are neglected in this model. 

These equations are evolved numerically in a 3-D spherical domain using 4th order integration method until a quasi-streamer belt is formed \citep[see,][for details]{Ofman2015}. The ranges for spatial coordinates are $1R_s \leq r \leq 7R_s$, $0 <\theta < \pi$ and $0 < \phi < 2\pi$, with a small circular region around the poles removed to avoid singularity of the coordinate system. The computation started with a tilted dipole (with a tilt of $10^o$) coronal magnetic field model \citep{Ofman2015}. The following boundary conditions are implemented in the simulation. At the inner boundary $r=1R_s$, the magnetic field components are fixed (line-tied), and the values of density, temperature, and velocity components at the boundary grid cells are extrapolated from the first interior cells with zero gradients,  approximating outgoing characteristics \citep{Steinolfson1976}. At the outer boundary $r = 7 R_s$ open boundary conditions are set for all variables. The outer boundary distance is chosen to satisfy supersonic slow solar wind transition. The presented results were obtained with a uniform resolution of $160\times128\times508$ grid cells in the $\theta$, $\phi$ and $r$ domains, respectively. The results were also tested for convergence with higher resolutions (albeit shorter duration) run and no significant changes were observed with higher resolution solutions. 

\section{Numerical Results}
\label{sec:numresult}
\begin{figure}
\begin{center}
\includegraphics{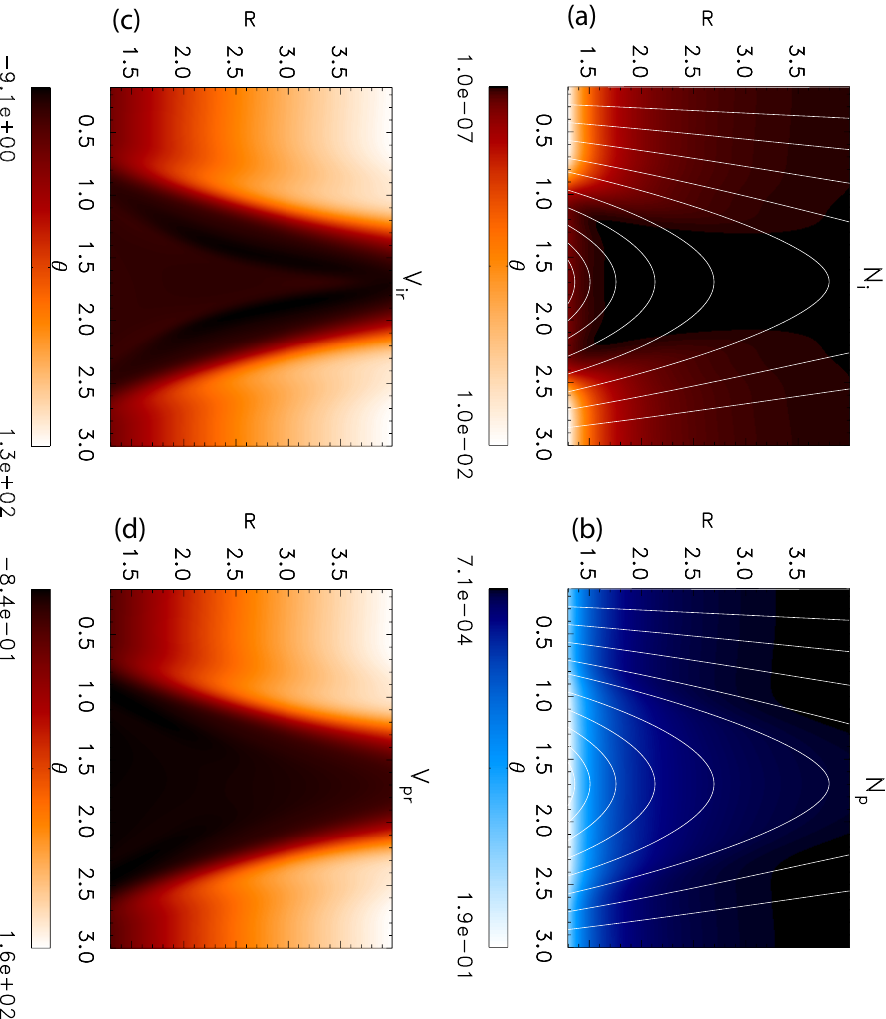}
\caption{The results of the 3D three-fluid model of the tilted streamer belt shown in the $r-\theta$ plane at longitude $\phi=0.76$ Rad. The distance is in $R_s$ and angles are in Rad. The densities are normalized in terms of $n_{e0}=5\times 10^8$ cm$^{-3}$. The while lines mark several fieldlines. (a) The normalized density of the $He^{++}$ ions, (b) the normalized proton density (blue), (c) the $He^{++}$ radial velocity in km s$^{-1}$, (d) the proton radial velocity in km s$^{-1}$.}
\label{fig:ninpvivp}
\end{center}
\end{figure}

Here, we present the results of our 3D three-fluid model of streamer belt with $He^{++}$ for the titled dipole initial state and the parameters given in the previous section. The solutions are obtained by running the model to quasi-steady state, where the initial tilted dipole field evolves to a streamer belt solution (as in  \citet{Ofman2015} but with $O^{5+}$). 
In Figure~\ref{fig:ninpvivp} we show the section of the streamer belt in the $r$-$\theta$ plane cut at $\phi=0.76$ Rad in the radial extent $1.3R_s\leq r \leq 4R_s$, similar to the radial extent covered by HERSCHEL observations. The while lines mark several closed and open fieldlines in this plane. It is evident that the slow solar wind outflow velocity $v_r$ is significant outside the streamer core, and velocities are small inside the streamer core for both, protons and $He^{++}$. The corresponding density structure of the $He^{++}$ dips significantly at the closed-field core of the streamer, and peaks in the open-field streamer flanks. The proton density structure shows the opposite latitudinal dependence, and peaks in the streamer core. This streamers density structure is in good qualitative agreement with observation in the region where $A_{He}$ was determined, shown in Figure~\ref{fig:herschel}.

\begin{figure}
\begin{center}
\includegraphics[angle=270,width=7in]{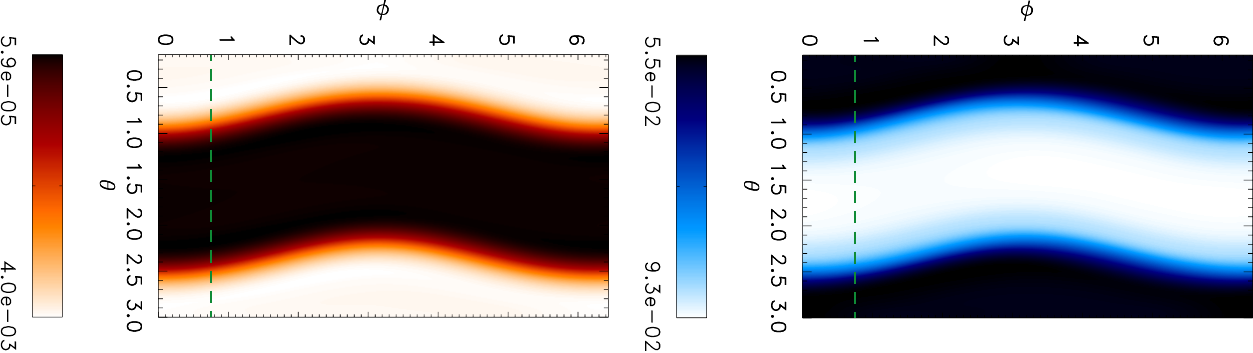}
\caption{The results of the 3D three-fluid model of the tilted streamer belt shown in the $\theta-\phi$ [Rad] plane at the height of $r=1.5R_s$. The densities are normalized in terms of $n_{e0}=5\times 10^8$ cm$^{-3}$.  Left panel: the normalized proton density. Right panel: the normalized density of the $He^{++}$ ions. The dashed green line marks the longitude of the $\theta-r$ plane cuts in Figure~\ref{fig:ninpvivp}.}
\label{fig:ninpvivp_xy}
\end{center}
\end{figure}

In Figure~\ref{fig:ninpvivp_xy} we show the density structure of the tilted streamer belt in the $\phi-\theta$ plane (i.e., analogous to a Carrington map but for density) at the height of $r=1.5R_s$. The apparent sinusoidal structure of the streamer belt is due to the $10^o$ tilt projected onto the plane, and the higher proton density in the streamer core, anti-correlated with the $He^{++}$ density is evident in the tilted streamer belt structure. The modeled tilted streamer belt structure is consistent with the Carrington map of the streamer belt for the solar minimum conditions during HERSCHEL observations shown in Figure~\ref{fig:pfss}.

\begin{figure}
\begin{center}
\includegraphics[angle=90,width=6.5in]{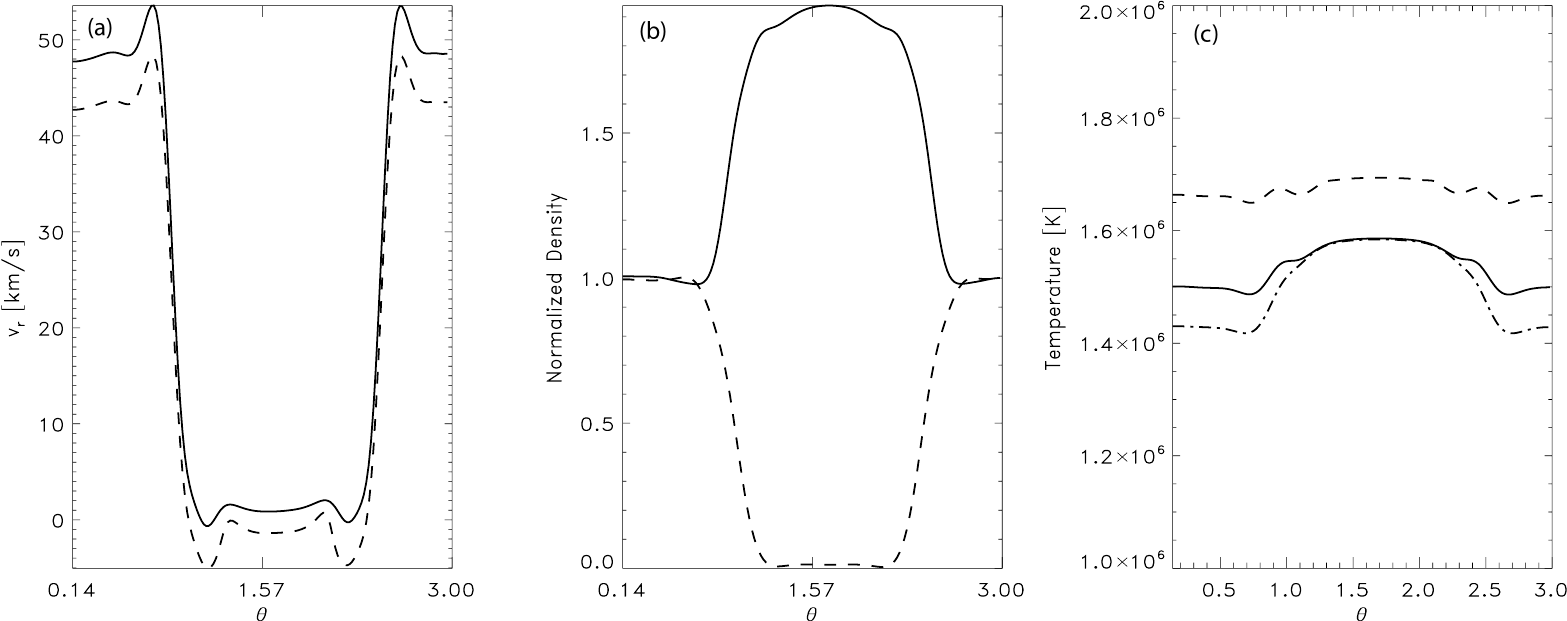}
\caption{The cross section cut of the variables at $r=1.75R_s$ for the model run shown in Figure~\ref{fig:ninpvivp} (solid: protons, dashes: $He^{++}$ ions, dot-dashes: electrons). (a)  The outflow velocities $V_r$, (b) the densities normalized by the values at maximal $\theta$. (c) Proton,  $He^{++}$, and electron temperatures.}
\label{fig:vrtheta}
\end{center}
\end{figure}

Figure~\ref{fig:vrtheta} is devoted to the latitudinal dependence of the variables across the streamer belt shown at a height of $1.75R_s$. The radial velocities $v_r$ of protons (solid line) and $He^{++}$ (dashes) are shown in Figure~\ref{fig:vrtheta}a. The  outflow velocities of the two ions are close due to the effects of the Coulomb friction terms with magnitude that is proportional to the velocity difference of the electrons, protons, and He$^{++}$ ions (the  $F_k$ terms in the momentum equation). The effect of the Coulomb friction is to reduce the differential flow of the three fluids. The modeled solar wind velocity outside the streamer is consistent with observational values \cite{Abbo2010,Ofman2011}. 
It is evident that inside the streamer belt the velocities are very low so that the plasma can be considered nearly static.
The small downflow velocity of $He^{++}$ is associated with gravitational settling. The anti-correlated normalized density structures of protons and $He^{++}$ ions are shown in Figure~\ref{fig:vrtheta}b. The densities are normalized with the corresponding densities at $\theta=3$ Rad. The temperatures of protons, $He^{++}$ ions, and electrons obtained from the three fluid model are shown in Figure~\ref{fig:vrtheta}c. It is evident that the $He^{++}$ ions remain hotter than the other fluids due to the overall effect of the ion heating term. However, in the core of the quiescent streamer the $He^{++}$ temperature is larger than in the open-field region, due to the thermal energy exchange and cooling by the protons and electrons. It is also evident that in the core the quiescent streamer the electron and proton temperature are nearly identical due to the thermal energy exchange terms that tend to minimize the temperature difference between the species when collision are significant.
\section{Discussion and Conclusions}\label{sec:disc}
The variability of helium abundance in the solar wind and the solar corona is a puzzling phenomenon  that provides clues on the solar cycle, type of the solar wind (fast or slow), and possibly on the acceleration processes and on the origins of the solar wind \citep[e.g.,][]{Alterman2019}. However, the processes that lead to the variability of helium in the corona are not well understood so far. The observation of UV helium emission is not routinely available in coronal streamers.  The only available observations so far, provided UV imaging of streamers in both, H~I and He~II emission using the HERSHEL rocket experiment reported recently by \citet{Moses2020}. It is evident from the imaging data of the inner corona that the emission of hydrogen and helium is overall anti-correlated, suggesting that different processes shape the helium abundance in streamers than hydrogen, as predicted by past three-fluid models. Moreover, the potential magnetic field extrapolation double streamer magnetic structures shown in \citet{Moses2020} does note account well for the gravitational settling of $He^{++}$ in a quiescent streamer core at the equatorial region, nor with the overall magnetic structure of an equatorial streamer belt at solar minimum. 

We use the 3D three-fluid model with electron, proton, and $He^{++}$ ion fluids to compute the tilted streamer belt structure in the inner corona in the region observed by HERSHEL instruments. We find good qualitative agreement of the observed streamer structure of protons and $He^{++}$ ions, that also agrees with previous more simplified 2.5D three-fluid models of the slow solar wind in coronal streamers, as well as 3D model with $O^{5+}$ ion as the third fluid with depleted equatorial streamer belt core helium abundance and enhanced helium abundance in the open field regions. Specifically, \citet{Moses2020} computed the helium relative abundance shown in their Figure~3 from the ratio of the H~I and He~II emissions, and we find qualitative agreement with our three-fluid modeling results.

The present model shows that the main processes that shape the distinct helium streamer structure and slow solar wind helium abundance are gravitational settling of $He^{++}$ in the core of the quiescent streamer closed-field region as well as collisional and thermal coupling  of $He^{++}$ ions to the electrons and protons. The increase $He^{++}$ relative abundance in open field region is produced by the velocity-dependent Coulomb friction process with the out-flowing electrons and protons, that tend to minimize differences between the fluids and as a result 'drags-out' the heavier $He^{++}$ ions. It is interesting to note that the observed depleted $He^{++}$, and the enhanced protons abundance in the streamer belt core is is captured by the idealized tilted streamer belt structure appropriate for solar minimum conditions, and does not require a complex multiple streamer configuration suggested in the past \citep{Noci1997,Moses2020}. However, reproducing finer details of the observation would indeed require more realistic magnetic field structure in the 3D multi-fluid model, as well as detailed emission calculations and line-of-sight integration, and these are left for future studies. Our modeling results that find very low helium abundance in the core of streamer are in agreement with recent in-situ observations of low $A_{He}$ at 1AU traced back to coronal streamer structures \citep{Yogesh2024}. The present results extend the previous three-fluid modeling studies with $He^{++}$ to more realistic 3D tilted streamer belt structure, appropriate for solar minimum conditions. 

\section{Acknowledgements} We would like to acknowledge discussions with Dr. Lucia Abbo. L.O. and Y. acknowledge support by NSF grant AGS-2300961, L.O. acknowledges support by NASA LWS grant 80NSSC20K0648. Resources supporting this work were provided by the NASA High-End Computing (HEC) Program through the NASA Advanced Supercomputing (NAS) Division at Ames Research Center. We would like to acknowledge high-performance computing support from Cheyenne (doi:10.5065/D6RX99HX) provided by NCAR's Computational and Information Systems Laboratory, sponsored by the National Science Foundation.


\end{document}